\documentclass{aa520}
\usepackage{txfonts}
\usepackage {graphicx}
\usepackage{placeins}
\usepackage {natbib}
\bibpunct{(}{)}{;}{a}{}{,}
\usepackage{longtable,array,tabularx}
\begin{document}
\title{NIR Observations of the QSO \object{3C 48} Host Galaxy
\thanks{Based on observations with the VLT; proposal number 67.B-0019}}
\author{J. Zuther\inst{1}, A. Eckart\inst{1}, J. Scharw\"achter\inst{1}, M. Krips\inst{1,2} \and C. Straubmeier\inst{1}}
\institute{1. Physikalisches Institut, Universit\"at zu K\"oln,
Z\"ulpicher Str. 77, 50937 K\"oln 
\and IRAM, 300, rue de la Piscine, Domaine Universitaire, 38406 Saint
Martin d'Heres }
\offprints{J. Zuther, \email{zuther@ph1.uni-koeln.de}}
\date{Received  / Accepted }
\abstract{In this paper we present new near infrared (NIR) imaging and
spectroscopic data of the quasar \object{3C 48} and its host galaxy. The data
were obtained with the ESO-VLT camera ISAAC.
We report the first detection of the apparent second nucleus 3C 48A
about 1\arcsec NE of the bright QSO nucleus in the NIR bands $J$, $H$, and $Ks$. 
3C 48A is highly reddened with respect to the host, which could
be due to warm dust, heated by enhanced star formation or by
interstellar material intercepting the radio jet. In fact,
all colors on the host galaxy are reddened by several magnitudes 
of visual extinction. Imaging and initial spectroscopy also reveal a
stellar content of about 30\% to the overall QSO-light in the NIR. 
These results are important input parameters for future models of the stellar populations by taking extinction into account.
}
\authorrunning{J. Zuther et al.}
\maketitle

\section{Introduction}
Revealing the physical properties of quasi stellar objects (QSOs) and
their host galaxies is one key to understand how the overall galaxy
structure affects the extreme activity in active galactic
nuclei (AGN) and how in turn the QSO and its black hole influence the
surrounding galaxy structure. 

The host galaxy of the radio-loud quasar 3C 48 is an unusually large
and bright object with excess far-infrared emission, $L_\mathrm{FIR}=5\times
10^{12}L_\odot$ \citep{1985ApJ...295L..27N}, showing a one-sided radio
jet extending about 0.5\arcsec to the north, fanning out to the east to
about 1$''$ \citep{1991Natur.352..313W}. 
At its redshift of $z\approx 0.3695$ \citep{1984ApJ...281..535B}
\object{3C 48} is one rare member of the class of compact steep spectrum (CSS)
radio sources with a radio-spectral index $\alpha\sim 0.9$
\citep{2001A&A...373..381T}. 
The quasar is rich in molecular gas, $M_{\mathrm{H}_2}\approx 2.7\times 10^{10}
M_\odot$, as CO(1-0) observations show 
\citep{1993ApJ...415L..75S,1997A&A...322..427W}.

There is also morphological evidence for a recent merger event, like a
possible double nucleus and a tidal tail extending several arcseconds
to the northwest \citep[~and references therein]{1991AJ....102..488S,
1999A&A...343..407C, 2000ApJ...528..201C}. But it is still unclear,
whether the brightness peak $\sim 1''$ northeast (NE) of the QSO is 
really the nucleus of the companion galaxy, as it could also be a region
of interaction of the radio jet with the interstellar medium of the
host galaxy, inducing star formation \citep{2000ApJ...528..201C} or even
both. \cite{2000ApJ...528..201C} found unusually high stellar velocity
dispersions as well as very young stellar populations, possibly due to
the interaction of the jet with the ambient medium \citep[see
also][]{1999A&A...343..407C}. The merger morphology, its high FIR
luminosity, and its large content of molecular gas put 3C 48 in an
evolutionary scheme envisioned by \citet{1988ApJ...325...74S}. It
fits nicely as an transition object between ultra-luminous infrared
galaxies (ULIRGs) and pure QSOs, being still dominated by its FIR
excess but also displaying QSO features like its luminosity and rather
flat spectral energy distribution.

The 3C 48 nucleus is an ideal target for high-resolution near infrared
studies of AGN and their hosts, in order to study the influence of the
nuclear activity as well as the ongoing merger on the host environment.
3C 48 also offers access to a unique set of diagnostic lines
(Pa$\beta$, Pa$\gamma$, \ion{Si}{I}, CO(6-3) and see Table
\ref{3C48DiagnosticLines}) to study stellar populations, effects of
extinction, varying metallicity and initial mass function. 

A major advantage of the NIR is that contamination due to scattered
light is much smaller than in the visible. Previous work did not take
extinction due to molecular material into account, which must play a
role in the dusty and gas rich host galaxy. 
Therefore it is likely that the patchy velocity and age distribution 
found by \cite{2000ApJ...528..201C} may be in part due to variations
in extinction.

Section 2 of this article describes details of our new NIR
observations, the data reduction procedures, and calibration in both
imaging and spectroscopy. In Section 3 we present and discuss the
results of photometry of the nuclear region, the 3C 48A potential
second nucleus and the tidal tail, and we present the initial
spectroscopic results. Section 4 gives a final discussion and
summary.

\section{Observations and Data Reduction}

\begin{figure*}[ht]
\resizebox{\hsize}{!}{\includegraphics{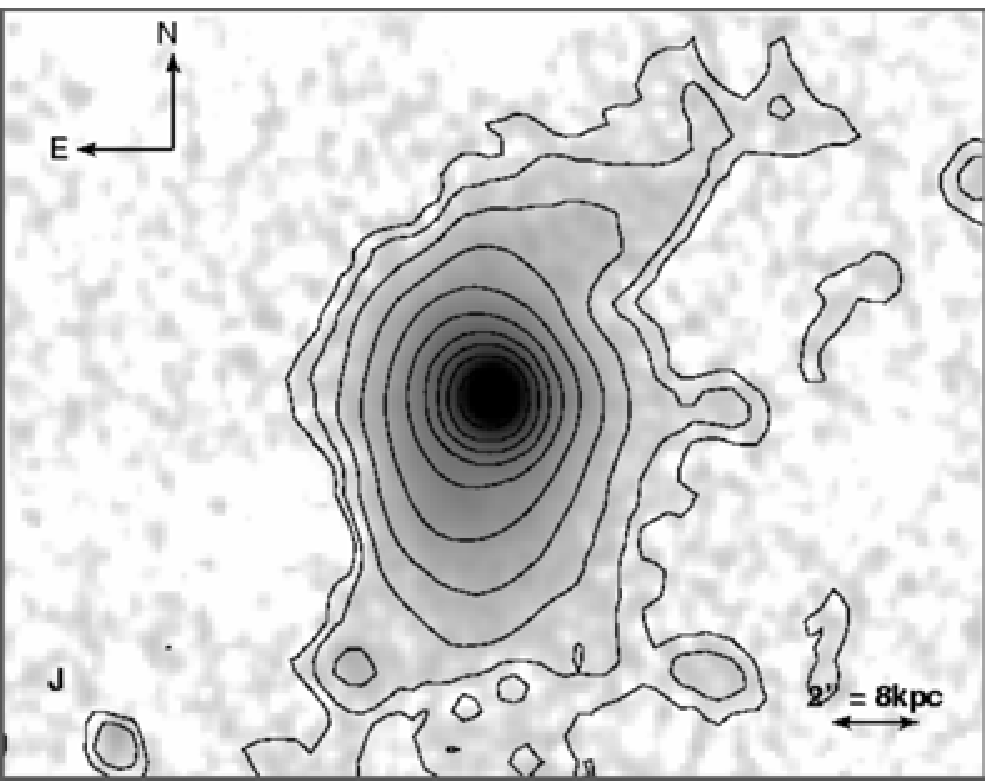}
\includegraphics{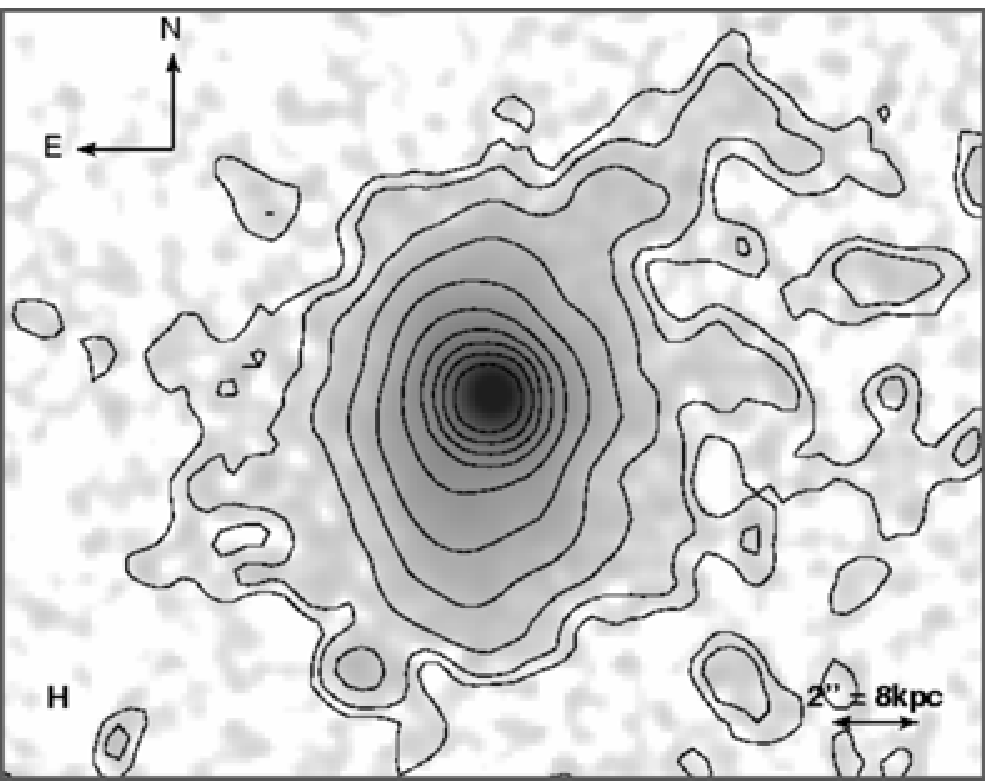}
\includegraphics{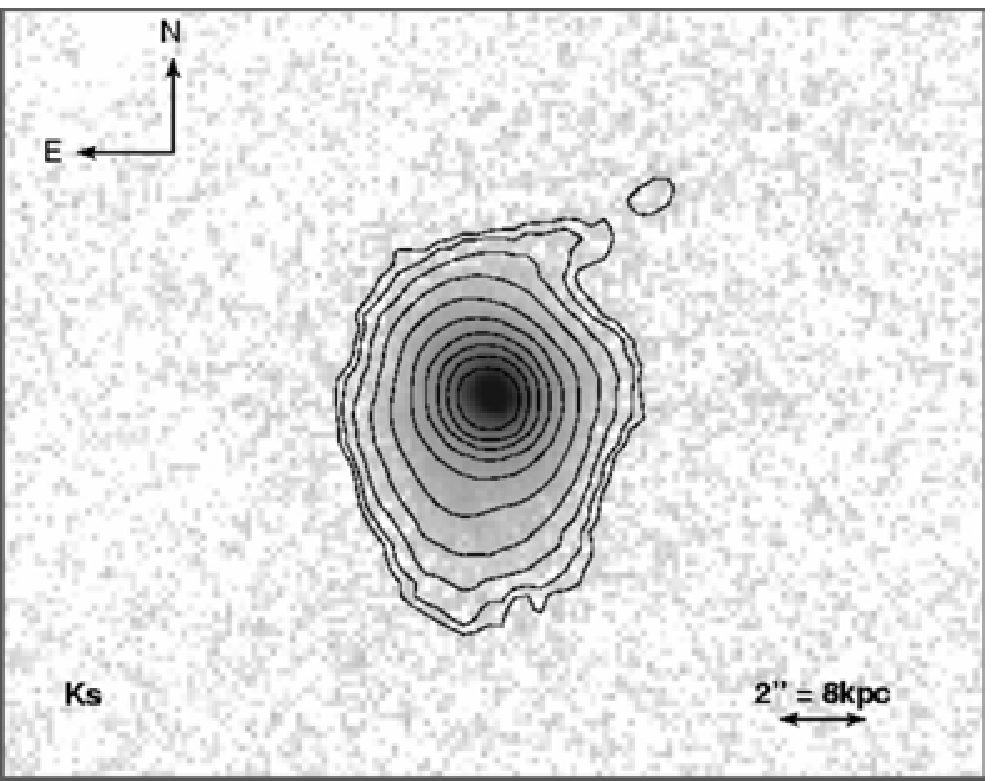}}
\caption{JHKs ISAAC images of
3C 48. The left panel displays the J, the middle panel the H
and the right panel the Ks image. All frames are smoothed with a
Gaussian to the same seeing (FWHM) of about 0.5\arcsec (given by Ks).}
\label{3c48jhk}
\end{figure*}

The data were acquired with the Infrared Spectrometer
and Array Camera (ISAAC, \citealt{1995SPIE.2475..262M}) mounted at the
Nasmyth B focus of Unit Telescope 1 (UT1, Antu) of the Very Large Telescope
(VLT; ESO, Chile).

\begin{table}[ht]
\caption{Journal of Imaging Observations. $t_{\mathrm{int}}$ is the integration time
in each band.}
\begin{center}
\begin{tabular}{r c c}
\hline
Date & Band & $t_{\mathrm{int}}$~[s]\\
\hline
30.07.01 & $Ks$ & 720 s\\
03.08.01 & $J$  & 1680 s\\
03.08.01 & $H$  & 720 s\\
24.09.01 & $J$  & 1200 s\\
24.09.01 & $H$  & 1440 s\\
24.09.01 & $Ks$ & 360 s\\
\hline
\end{tabular}
\end{center}
\label{imJournal}
\end{table}

\begin{table}[ht]
\caption{Journal of Spectral Observations. $t_{\mathrm{int}}$ is the integration time
in each band.}
\begin{center}
\begin{tabular}{r c c}
\hline
Date & Band & $t_{\mathrm{int}}$~[s]\\
\hline
18.08.01 & $H$ & 960 s\\
20.09.01 & $H$ & 2400 s\\
20.09.01 & $K$ & 2400 s\\
\hline
\end{tabular}
\end{center}
\label{specJournal}
\end{table}

\subsection{Imaging}
The data cover the short wavelength range from 1.2 - 2.2~$\mu$m. 
Imaging was performed in the broad bands $J$, $H$ and $Ks$.
The 1024x1024 pixel Hawaii Rockwell array detector provides a pixel
scale of 0.1484\arcsec/pixel with a field of view of 152x152~
arcsec$^2$. Tables \ref{imJournal} and \ref{specJournal} list the
observational parameters. 
The data were reduced with IRAF using standard procedures. 
Considering our data, we concluded by visual inspection, that the
flat fields in all bands seem to be adequately flat with a variation
smaller than 3\% over the area of interest. The science object was
also moved across the array, so that these variations are reduced due
to the averaging of the individual exposures. Therefore no flat field
correction was applied. 
Successive object and sky observations were used to produce sky
subtracted images. Shifting all sky subtracted images with respect to
one reference image and a median averaging results in the images
displayed in Figure \ref{3c48jhk}.
All these resulting images were smoothed to a common seeing limited
resolution of 0.5\arcsec.
Calibration of our data relies on the observation of the three
standard stars FS10 \citep{2001MNRAS.325..563H}, S677D and
S279-F \citep{1998AJ....116.2475P}. 
Airmass-corrected zero-points, ZP$_c$, for the three bands, are
presented together with the used apertures in Table
\ref{3c48OverallMag}. The extinction coefficients $k$ are mean values
supplied by the ESO-Paranal web page.
An average ZP in each band was computed and is used for the 3C 48
photometry. The ZP errors were calculated to be of the order of
$\Delta J=\pm 0.03$ mag in $J$, $\Delta H=\pm 0.07$ mag in $H$ and $\Delta
Ks=\pm 0.05$ mag in $Ks$.

\subsection{Spectroscopy}
In order to study the nucleus and host of 3C 48 we used slit spectroscopy
to get spectra of a number of regions along the slit in bands $H$ and $K$.
The slit was positioned such, that it provided a north-south cut through the galaxy including part of the disk, the active nucleus, and the boundary region of the potential second nucleus.
As with imaging the sky was subtracted. Again, the flat fields are
flat to within better than 3\%, so that no correction was applied.
The individual 2-dimensional spectra were geometrically corrected.
After combining the shifted, double sky subtracted 2-dimensional
spectra, 1-dimensional spectra were extracted using standard IRAF
procedures. The integration time in both filters $H$ and $K$ was too short
(because of the proposal being B-rated, only about 1 hour and 0.6 hours respectively have been acquired) to reveal many spectral
features. These give upper limits for the $H$- and $K$-lines in emission
or absorption (see Section 3.2). 
We thus extracted only one spectrum per band from the nuclear region
of the quasar. The FWHM of the 3C 48 trace is $\sim 5$ pixels in $K$ and
$\sim 10$ pixels in $H$. This has to be compared with the FWHM of the
calibration star traces which are about 11 pixels.
While extracting the spectrum a sky is subtracted by sampling
a region above and beneath the trace. Using the IRAF task
\emph{telluric} we also removed most of the atmospheric lines.

The continua of the resulting spectra of 3C 48 are quite flat.
We used the flux\footnote{The flux is calculated from magnitudes via
$f^\star_\lambda=f_\lambda\times 10^{-0.4 m^\star_\lambda}$ where
$f^\star_\lambda$, in units of W~cm$^{-2}\ \mu$m$^{-1}$, is the science
object's flux, using $f_\lambda$ for a 0mag-star as given in Table 
\ref{fluxcalib}.} of the nuclear region from our NIR imaging data to
flux-calibrate the spectra. Using the facts of the flatness of
the spectra and the absence of strong line features, one can distribute
the imaging data flux, $f^\star_\lambda$ in W~cm$^{-2}\ \mu$m$^{-1}$,
over the filter-bandwidth\footnote{We assumed that the imaging and
spectral filters are identical to first order, because the ISAAC
spectroscopic filters are somewhat broader than the photometric
ones.}:

\begin{equation}
f'_\lambda=\frac{\mathrm{counts}}{\mathrm{sum~ of~ counts~ over~ filter~
bandwidth}}\times f_\lambda^\star 
\end{equation}
where $f'_\lambda$ in units of W~cm$^{-2}\ \mu$m$^{-1}$ is the new
y-axis . 

\begin{table}[ht]
\caption{Absolute flux calibration
of 3C 48 spectra. Flux densities $f_\lambda$ for a mag=0
star \citep[~and references
therein]{1999hia..book.....G}. $f_\lambda^\star$ is the flux from
imaging data.} 
\label{fluxcalib}
\begin{center}
\begin{tabular}{c c c c}
\hline
Band & $\lambda_\mathrm{eff}~[\mu m]$ & $f_\lambda$~[W~
cm$^{-2}\ \mu$m$^{-1}$] & $f^\star_\lambda$~[W~cm$^{-2}\ \mu$m$^{-1}$] \\
\hline
$H$  & 1.63 & $1.14\times 10^{-13}$ & $2.27\times 10^{-19}$\\
$K$ & 2.19 & $3.96\times 10^{-14}$ & $1.67\times 19^{-19}$\\
\hline
\end{tabular}
\end{center}
\end{table}
In the spectra there are very few features detectable from the overall
noise in both spectra. The major reason for this seems to be the very
short integration time. The $H$ spectrum has a RMS noise of about
$4\times 10^{-23}$~W~cm$^{-2}\ \mu$m$^{-1}$ and the $K$ spectrum one of
about $5\times 10^{-24}$~W~cm$^{-2}\ \mu$m$^{-1}$.  

\section{Results and Discussion}
\subsection{3C 48 Surface Photometry}
\label{surfacePhotometry}
To place the QSO and the host galaxy into a two-color diagram,
photometry was carried out at several positions on the galaxy in the
three bands $J$, $H$, and $Ks$. 
\paragraph{Overall Colors:}
Overall magnitudes of 3C 48 were determined using an aperture
of a diameter of 10\arcsec (= 68 pixels) centered on the
centroid of the brightness distribution.
In this case one arcsecond corresponds to a linear scale of 4~kpc.
These magnitudes are dominated by the unresolved active nucleus.
Additionally two smaller apertures with diameters of 5\arcsec 
(= 34 pixels) and 1\arcsec (= 7 pixels) respectively have been
applied, centered on the same position. A comparison allowed to
estimate the stellar contribution to the AGN luminosity (see below).

The IRAF task \emph{qphot} (centering and integrating the flux) has
been used to determine the flux in ADUs in these apertures in the
individual bands. Table \ref{3c48OverallMag} displays the resulting
magnitudes using the above determined zero-points.
The error is dominated by sky-noise, determined from the standard
deviation within 5x5 pixel boxes at several positions on the sky. This
yields a precision in magnitudes of about $\Delta \mathrm{m} \approx
\pm 0.1$ in the individual bands.

From these magnitudes the $H-K$ and $J-H$ colors for
the three apertures have been calculated (Table \ref{3c48OverallColor}).

\begin{figure}
\begin{center}
\resizebox{\hsize}{!}{\includegraphics{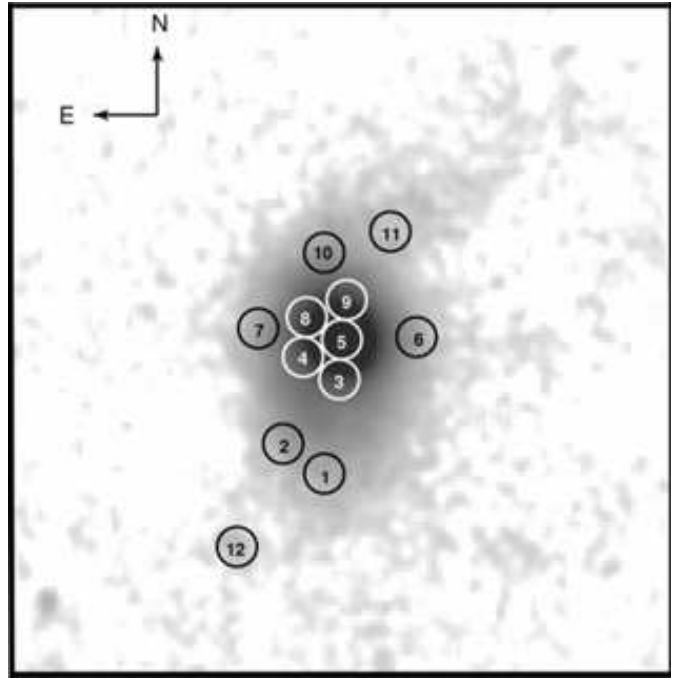}}
\caption{Positions of apertures with radii of 1.6\arcsec on 3C 48. 
The apertures are labeled with numbers.
The size of the map is 26$\times$26 arcsec$^2$ corresponding 
to 104$\times$104~pc$^2$.}
\label{3c48ManyAp}
\end{center}
\end{figure}

\paragraph{Colors of the Host Galaxy:}
In order to obtain information on the nature of the host galaxy, it is
necessary to carry out photometry at several positions on the
host. Using for example models for color evolution
\citep{1993ApJ...405..538B, 1997AJ....113..550H}, one can estimate the
age of the stellar content residing in the disk of the host
galaxy. With surface photometry one can additionally investigate the
importance of reddening due to (warm) dust. 
We therefore measured the flux at several positions across the galaxy,
which are indicated in Figure \ref{3c48ManyAp}. Altogether we choose
12 apertures of 1.2\arcsec (8 pixels) diameter at representative
positions. Three apertures are lying in the south and two in the tidal
tail region to the northwest. 
Five apertures lie around the active nucleus 
and cover the location of the potential second nucleus $\sim 1''$ to
the northeast. Aperture number 12 lies at the position within an
apparent counter tidal \citep{2000ApJ...528..201C} reaching out to
about 7\arcsec to the southeast (see Fig. \ref{3c48ManyAp}). In our NIR
images we do not see a tidal tail feature reaching out to about
8\farcs 5 as proposed in \cite{1999MNRAS.302L..39B}.

\begin{figure*}
\sidecaption
\includegraphics[width=12cm]{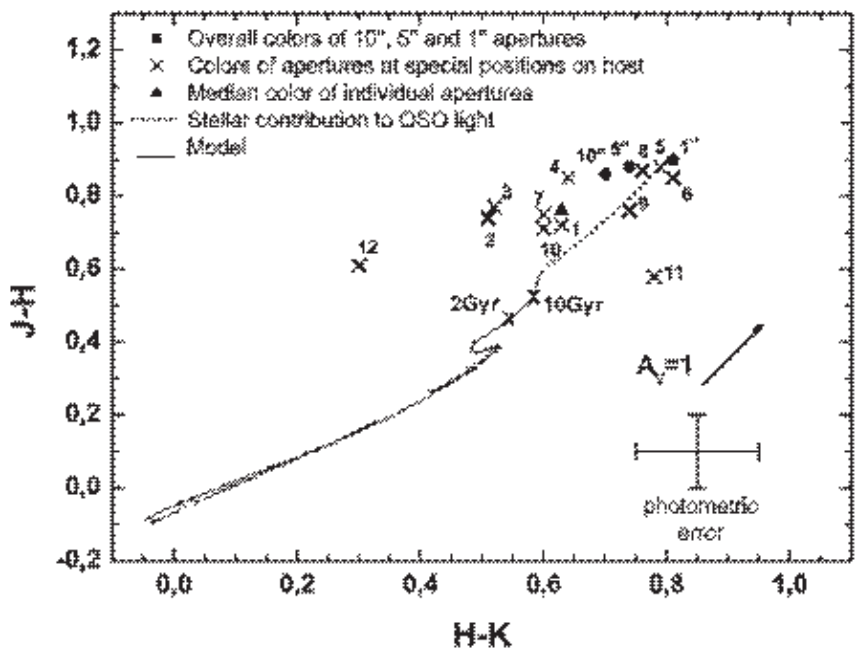}
\caption{
$J-H$ vs. $H-K$ two color diagram showing the overall colors (filled circles with aperture diameter given in arcsec) and colors derived from apertures on
distinct positions on the galaxy (crosses with aperture
number as label (see Fig. \ref{3c48ManyAp})). The median color of all
apertures without the nuclear one is shown as a triangle. The solid
line is an evolutionary model for colors
\citep{1997AJ....113..550H}. Evolution proceeds from the lower left
corner at 0~Gyr to the upper right corner at 20~Gyr. The initial
population is a 1~Gy starburst followed by passive evolution. The
dotted color line represents the stellar contribution to the QSO light
when increasing the aperture (cf Fig. \ref{3c48CoreSubtracted}). The
arrow indicates reddening in colors for a visual extinction of
$A_V=1$ \citep{1985ApJ...288..618R}.}
\label{3c48TwoColors}
\end{figure*}

\begin{figure*}
\sidecaption
\includegraphics[width=12cm]{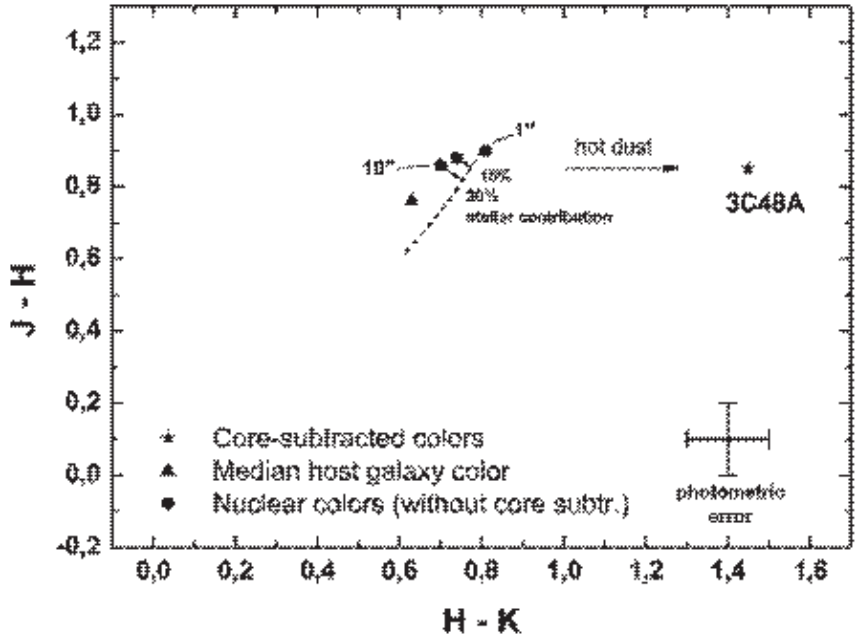}
\caption{
Two color diagram showing the NIR properties of the brightness-excess
region 3C 48A after subtraction of the active nucleus contribution
(stars) together with the median color of the host (triangle) and the
overall colors (dots) already shown in figure
\ref{3c48TwoColors}. Also shown is the stellar contribution to the 
overall color, when increasing the aperture (dotted line).}
\label{3c48CoreSubtracted}
\end{figure*}

\paragraph{$J-H/H-K$ two Color Diagram of 3C 48:}
The two-color diagram in Fig. \ref{3c48TwoColors} shows (1) the overall
colors, (2) the colors derived from the 12 distinct positions on
interesting locations in the 3C 48 host (as discussed above) and (3) a
model for color-evolution at a redshift of $z=0.35$ (of about the
distance to 3C 48) \citep{1997AJ....113..550H} and (4) the stellar
contribution to the light measured within an aperture centered on the
nucleus (see also Fig. \ref{3c48CoreSubtracted}). 

(1) \emph{Overall colors}. Considering only the overall colors (green
dots), there is a trend of redder colors in $H-K$ as well as in
$J-H$ for smaller apertures (10\arcsec down to 1\arcsec). This is in
agreement with the idea, that in a large aperture there is a
contribution of both the host galaxy (filled circles) and the active
nucleus. In a smaller aperture, in contrast there only remains the
nuclear contribution which is usually redder due to gas and dust
the active core is embedded in. The dotted line in the
two color diagram (Fig. \ref{3c48TwoColors}) presents the influence of
mixing the nuclear color, determined in the 1\arcsec aperture, with a
stellar contribution, taking typical stellar colors at the 10~Gyr
point\footnote{We have chosen this age, because the dominant,
underlying stellar population is of about 10~Gyr
\citep{2000ApJ...528..201C} and is also typical for 'inactive' stellar
dominated galaxies.} of the evolutionary curve and holding the flux in
the Ks-band constant, so that the total flux is conserved for the
start and end points. Starting with no stellar contribution at
the point of the 1\arcsec (4~kpc) aperture and arriving at 100\% stellar
contribution at the 10~Gyr color point, the data points of the two
larger apertures lie on this curve within the errors and correspond to
a stellar contribution of about 15\% for the 5\arcsec (20~kpc) aperture
and about 30\% for the 10\arcsec (40~kpc) aperture (see
Fig. \ref{3c48CoreSubtracted}). The stellar contribution originates in
the disk of the host galaxy. The 10\arcsec aperture covers 
essentially the whole galaxy. Therefore the stellar contribution of
the underlying host galaxy to the light from the active nucleus is of
the order of 30\% at 2.2~$\mu$m. A consistent check for the stellar
content can be the CO(6-3) absorption, a tracer for the stellar
content \citep{1993A&A...280..536O}. One has to note that for the
1\arcsec aperture for the nuclear region a stellar contribution still has
to be taken into account.

\begin{table}[ht]
\caption{Photometry of 3C 48 with three different apertures centered on
the QSO}
\label{3c48OverallMag}
\begin{center}

\begin{tabular}{@{}p{0.8cm}@{}p{1.1cm}@{}p{1.2cm}@{}p{0.8cm}@{}p{1.5cm}@{}p{1.5cm}@{}p{1.4cm}@{}}

\hline
 Band & DIT~[s] &  Airmass & k    & ZP$_\mathrm{c}$~[mag] & Diam.~[\arcsec]  & Mag\\
\hline
$J$    & 20       & 1.90     & 0.11 & 25.13        & 10          & 14.4$\pm 0.1$\\
     &          &          &      &              & 5           & 14.4\\
     &          &          &      &              & 1           & 15.1\\  
\hline
$H$    & 15       & 2.01     & 0.06 & 24.77        & 10          & 13.5$\pm 0.1$\\
     &          &          &      &              & 5           & 13.6\\
     &          &          &      &              & 1           & 14.3\\  
\hline
$Ks$   & 10       & 2.01     & 0.07 & 24.26        & 10          & 12.8$\pm 0.1$\\
     &          &          &      &              & 5           & 12.8\\
     &          &          &      &              & 1           & 13.4\\  
\hline
\end{tabular}
\end{center}
\end{table}

\begin{table}
\caption{Colors for individual apertures centered on the QSO}
\label{3c48OverallColor}
\begin{center}
\begin{tabular}{c c c}
\hline
Diameter~[\arcsec] & $H-K$ & $J-H$\\
\hline
10 & 0.7$\pm 0.1$ & 0.9$\pm 0.1$\\
5 & 0.8$\pm 0.1$ & 0.9$\pm 0.1$\\
1& 0.8$\pm 0.1$ & 0.9$\pm 0.1$\\
\hline
\end{tabular}
\end{center}
\end{table}

(2) \emph{Colors from single apertures}. Colors derived from
measurements of aperture 12 are unusually blue compared to the colors
derived from other apertures on the host galaxy. Its colors are more
typical for normal galaxies at redshifts close to zero
\citep{1978ApJ...220...75F}. This suggests that 
this feature ($\sim 6.5''$ to the southeast of the quasar nucleus),
while being aligned with the apparent counter-tidal tail to the
southeast, in fact is a foreground galaxy. This contravenes with the
redshift derived in \cite{2000ApJ...528..201C} ($z\approx 0.8112$). As
both results assume that this feature is not associated with 3C 48,
further investigation is necessary.
Furthermore recent multi-particle simulations \citep{scharw} indicate,
that starting from an Antennae-like setup
\citep[e.g.~][]{1972ApJ...178..623T} one can arrive at a 3C 48
look-alike due to different projection parameters. In the case of
\cite{scharw}, the counter-tidal tail is extending from the SW towards
a NE direction in front of the galaxy host. Based on their HST images
\cite{2000ApJ...528..201C} already speculated on the presence of a
tidal tail in a SE-SW direction. This extension of the host towards
the SW is also visible in our J- and H-band maps
(Fig. \ref{3c48jhk}). But it could also belong to the tidal tail
resulting from the multi particle simulations. 

(2) and (3) \emph{Color-evolution model}. The two crosses on the evolutionary curve in Fig. \ref{3c48TwoColors} show the loci of stellar populations at an age of 2 and 10~Gyr, respectively. 
Noticing that the color of the upper right end of the model curve
corresponds to an age of 20~Gyr for the stellar population, a striking
result is that all derived colors are considerably reddened in
comparison to the model shown. Indicated in the figure is also the
effect of reddening for a visual extinction of $A_V=1$ (as discussed in
\citealt{1999hia..book.....G}). This indicates, that extinction
might play an important role in deriving the physical properties of
the host galaxy. As already mentioned, \cite{2000ApJ...528..201C} did
not take extinction into account in their calculations. They state
that the underlying old population has an age of about 10~Gyr. Taking
the median color of all our measurements on the host 
\begin{eqnarray*}
\mathrm{Median}(H-K)&=& 0.6\\
\mathrm{Median}(J-H)&=& 0.8
\end{eqnarray*}
we find that within the errors and including a visual extinction of
$A_V=1$ \citep{1985ApJ...288..618R} this result is consistent with an
age of 10 Gyr. But we also sample regions of active star formation (as
indicated by \citealt{2000ApJ...528..201C}), i.e. regions with much 
younger stellar components. 
The authors find e.g. a young starburst population within the tidal tail
at the position of our aperture 11. The $J-H$ color is about 0.25~mag
bluer than the median color of the host galaxy.
Its position is shifted in $H-K$ by about 0.2~mag to the red
with respect to the evolutionary track shown in Fig. \ref{3c48TwoColors}. 
This indicates a younger stellar population which
may be associated with warm dust that contributes to the red $H-K$ color in
emission. 
\cite{2000ApJ...528..201C} quote an age of the corresponding stellar
population of about 10~Myrs mixed with an underlying old stellar
population of about 10~Gyr. Starburst populations with those ages lie
at the lower (bluer) end of the color-evolution curve in the two color
diagram. 
The presence of warm dust and the location of the aperture in the two
color diagram, however, may indicate that a considerable reddening is
at work. In this case the optical radiation would only be indicative
of the properties of a fraction of the underlying stellar population.

(4) \emph{The circum-nuclear region}. 
The flux from central regions of
the host galaxy is still dominated by emission from the active
nucleus. In order to get a more detailed view on the underlying host
we subtracted the contribution of the active nucleus. Here we used the
fact that the QSO-nucleus is unresolved, i.e. star like in extent. 
We have done the subtraction by taking a star within a distance of
1\arcmin to 3C 48 shifting its centroid with subpixel-accuracy onto the
position of the 3C 48-centroid and subtracting it scaled to the flux
of the active nucleus. One has to take care that no 'holes' appear in
the galaxy after this process. Ideally this should give a smooth
brightness distribution in the central region after subtraction and
primarily the host contribution should remain.

\begin{table}[ht]
\caption{Nuclear Flux Contribution}
\label{3c48NuclearContribution}
\begin{center}
\begin{tabular}{c c}
\hline
Band & Mag  \\
\hline
$J$    & $15.1\pm 0.1$\\
$H$    & $14.2\pm 0.1$\\
$Ks$    &$13.5\pm 0.1$\\
\hline
\end{tabular}
\end{center}
\end{table}

\begin{figure*}[ht]
\resizebox{\hsize}{!}{\includegraphics{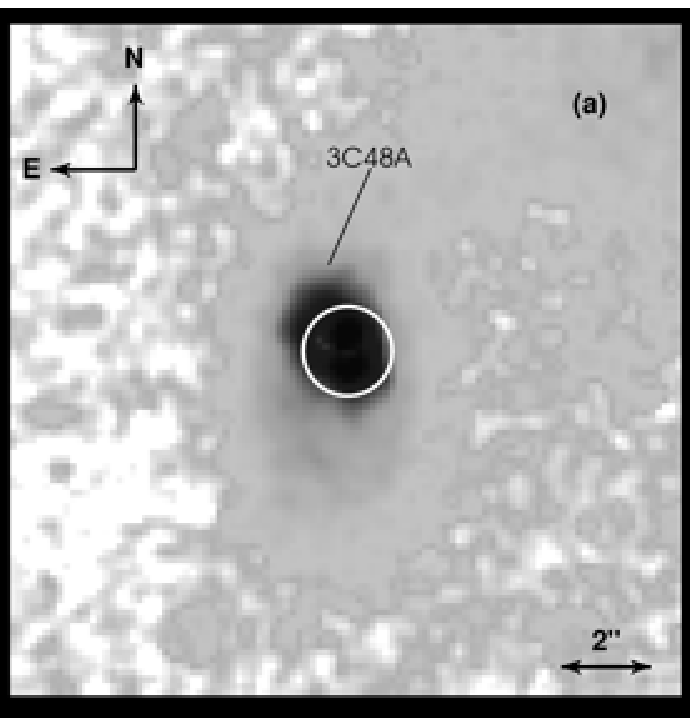}
\includegraphics{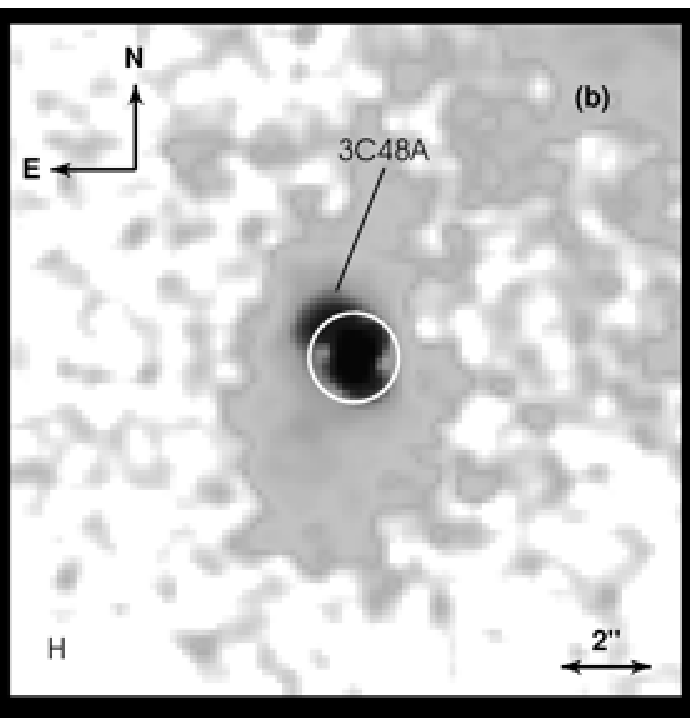}
\includegraphics{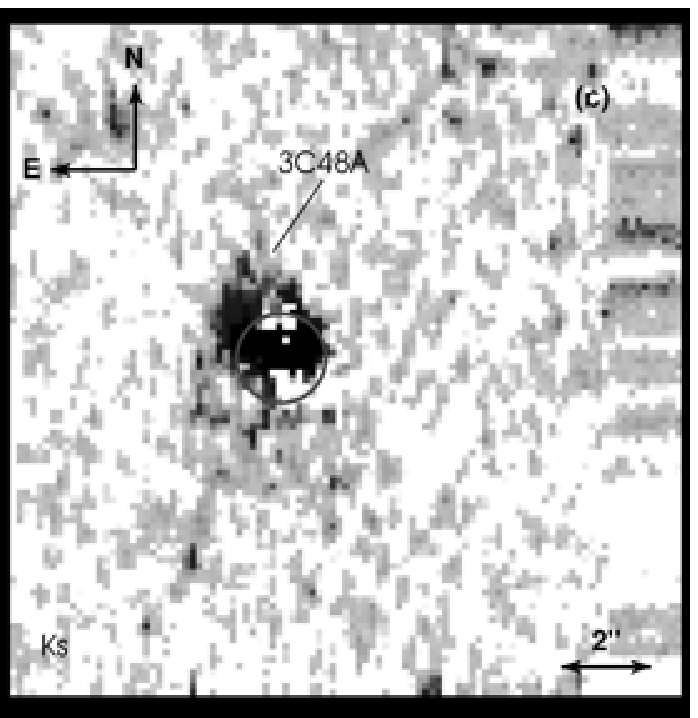}}
\caption{$JHK$ images with subtracted
nucleus (see text for details). (a) $J$, (b) $H$ and (c)
$Ks$. The circles of $1''$ radius centered on the active nucleus indicate
the location of 3C 48A northeast of the QSO. In all three bands an
additional flux component at the position of 3C 48A becomes visible
after the subtraction of the QSO nucleus.}
\label{3c48CoreSubtractedImages}
\end{figure*}

Figure \ref{3c48CoreSubtractedImages} displays the core-subtracted images
in the $J$-, $H$- and $K$-bands and Table \ref{3c48NuclearContribution}
lists the nuclear contributions which have been subtracted. After the
subtraction there remains no really smooth brightness distribution but
certain residuals. This is most probable due to psf-mismatching. It
would be favorable to take a composite psf by measuring several stars,
giving a more realistic sampling of the point-spread function,
impossible in this case due to the restricted FOV of the science
exposures. But a brightness excess region about 1\arcsec northeast of the
QSO is clearly visible in the residual images as indicated in Figure
\ref{3c48CoreSubtractedImages}.
\begin{table}[ht]
\caption{Parameters of Gaussian Profiles for 3C 48A. $\Delta$mag$_\mathrm{Peak}$
is the difference in magnitudes between the host contribution and host plus
3C 48A contribution.}
\label{3c48AGaussian}
\begin{center}
\begin{tabular}{c c c}
\hline
Band & FWHM~[\arcsec] & $\Delta$mag$_\mathrm{Peak}$\\
\hline
J    & 2.1           & 0.42\\
H    & 2.1           & 0.49\\
Ks   & 1.2           & 2.84\\
\hline
\end{tabular}
\end{center}
\end{table}
Its peak position in all three bands is $\sim 1''$ north-east of the
QSO-location, having a position angle of about 40$^\circ$ (as measured
from north to east). This feature is most prominent in the Ks image
whereas it gets fainter in the $H$ and $J$ images.

In order to estimate the amount of magnitudes with which 3C 48A
exceeds the underlying host galaxy emission, we fitted a Gaussian to
this emission region. After subtraction of a Gaussian with this
scaling we recognized a circular symmetric residual emission from
the central region as expected. On the opposite side of 3C 48 at the
same distance as 3C 48A we determined the flux value of the host
galaxy. The difference between these two values then gives an estimate
of the 3C 48A contribution presented in Table \ref{3c48AGaussian}.

$J-H$ and $H-K$ colors were determined by measuring the flux in $J$, $H$
and $Ks$ in apertures of the size of the FWHM of the Gaussian profile.
Figure \ref{3c48CoreSubtracted} displays the data in a two color
diagram. The determined colors of 3C 48A have to be taken with care,
because of the psf mismatch mentioned above. Additionally, the $Ks$ band
image is quite noisy which makes the measurement process more
complicated. Together this leads to a high degree of freedom in
choosing an appropriate Gaussian fit of this region. The apertures
have been chosen to be of the sizes listed in Table
\ref{3c48AGaussian}, i.e. rather small, in order to reduce the impact
of the imperfect QSO-subtraction process, which did not
result in a smooth brightness distribution and produced some
'holes'.

Figure \ref{3c48CoreSubtracted} shows that the 3C 48A area is heavily
reddened in the $H-K$ direction. This indicates the presence of hot dust
which usually shifts colors exclusively in the $H-K$ direction as it
influences predominantly the $K$ band \citep{1985MNRAS.214..429G}. 
\begin{figure}
\resizebox{\hsize}{!}{\includegraphics{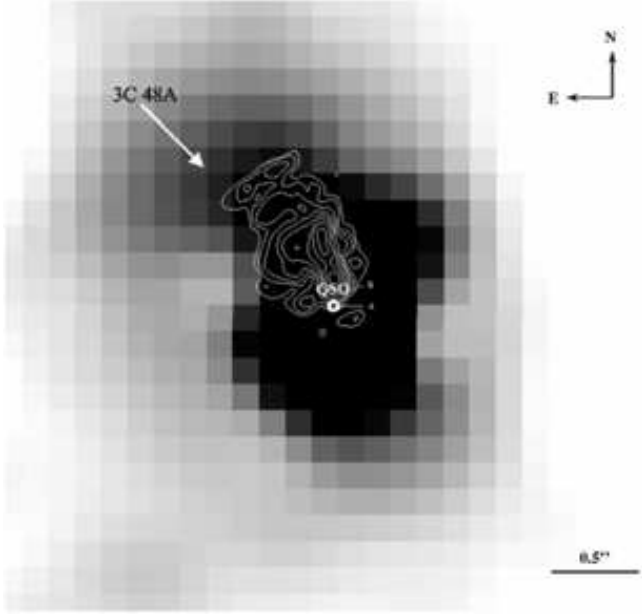}}
\caption{Overlay of the central region of 3C 48 in the $H$-band (centered on the QSO) with a 1.66~GHz radio continuum map from \citet{1991Natur.352..313W}, showing the expansion of the radio jet towards the location of 3C 48A.
}
\label{3c48Merlin}
\end{figure}

An overlay of the NIR images with a radio continuum map of Merlin and European VLBI data at 1.66~GHz \citep{1991Natur.352..313W} shows an apparent alignment of the radio jet and the position of 3C 48A (Fig. \ref{3c48Merlin}). Such an alignment was also found by \citet{1991AJ....102..488S}. Together with the results of \citet{scharw} there is still no definite answer on the origin of the excess brightness of 3C 48A. Probably it is a combination of both, a second nucleus and a jet-ISM interaction.

\subsection{The $H$-band Spectrum}
\begin{figure}
\resizebox{\hsize}{!}{\includegraphics{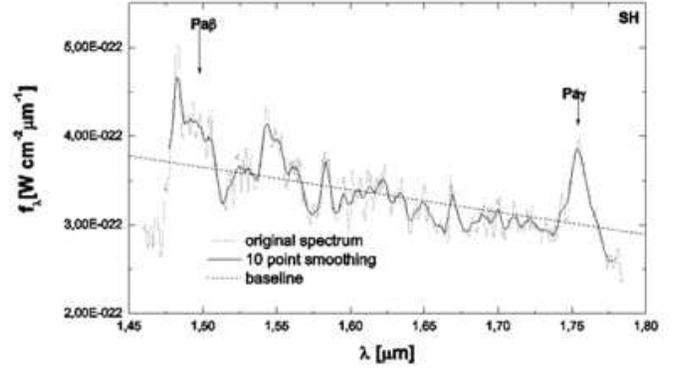}}
\caption{H band spectrum of the nuclear region of 3C 48. The dotted line is the original spectrum, whereas the solid line shows the spectrum after a 10 point smoothing. Indicated are the redshifted positions of the Pa$\beta$ and Pa$\gamma$ lines.
}
\label{3c48SHSpec}
\end{figure}

The most prominent features in the $H$ spectrum (Fig.~\ref{3c48SHSpec}) are the
Pa$\beta\ \lambda$17561 and Pa$\gamma\ \lambda$14985 lines (cf. Table
\ref{3C48DiagnosticLines}).
\begin{table}
\centering
\caption{NIR diagnostic Lines in 3C 48}
\begin{tabular}{c c c}
\hline
Line   & $\lambda_\mathrm{rest}$~[$\mu$m] & $\lambda_\mathrm{obs}$~[$\mu$m]\\
\hline
\multicolumn{3}{c}{$H$-band}\\
Pa$\beta$  & 1.2818 & 1.7561\\
Pa$\gamma$ & 1.0938 & 1.4985\\
\multicolumn{3}{c}{$K$-band}\\
CO(6-3)    & 1.6185 & 2.2173\\
\ion{Si}{I}        & 1.5892 & 2.1772\\
H$_2$(1-0) S(7) & 1.7475 & 2.3941\\
\ion{Fe}{II}       & 1.5335 & 2.1009\\
\ion{Fe}{II}       & 1.6435 & 2.2516\\
\ion{Fe}{II}       & 1.7449 & 2.3905\\
\hline
\end{tabular}
\label{3C48DiagnosticLines}
\end{table}

The baseline is subtracted using a straight line. A Gaussian fit to the Pa$\beta$ and Pa$\gamma$ lines gives line fluxes of
\begin{eqnarray}
f_{Pa\beta} &=&1.0\times 10^{-24}\ (\pm 20\%)\ \mathrm{W~cm}^{-2}\\
f_{Pa\gamma}&=&5.6\times 10^{-25}\ (\pm 20\%)\ \mathrm{W~cm}^{-2}
\end{eqnarray}

The line ratio Pa$\beta$$/$Pa$\gamma$ can then be calculated:
\begin{equation}
\frac{\mathrm{Pa}\beta}{\mathrm{Pa}\gamma}=1.8(\pm 30\%)
\end{equation}
When using line ratios for {\it Menzel's case B} (see
e.g. \citeauthor{1999hia..book.....G} \citeyear{1999hia..book.....G}
or~ \citeauthor{osterbrock89:_astrop_gaseous_nebul_activ_galac_nuclei}
\citeyear{osterbrock89:_astrop_gaseous_nebul_activ_galac_nuclei}),
this result is consistent within its errors with temperatures of
$T=10^4$~K and electron densities of $\rho=10^4$~cm$^{-3}$, which are
typically found for the atomic interstellar medium in QSO nuclei.

\subsection{The $K$-band Spectrum}
\begin{figure}
\resizebox{\hsize}{!}{\includegraphics{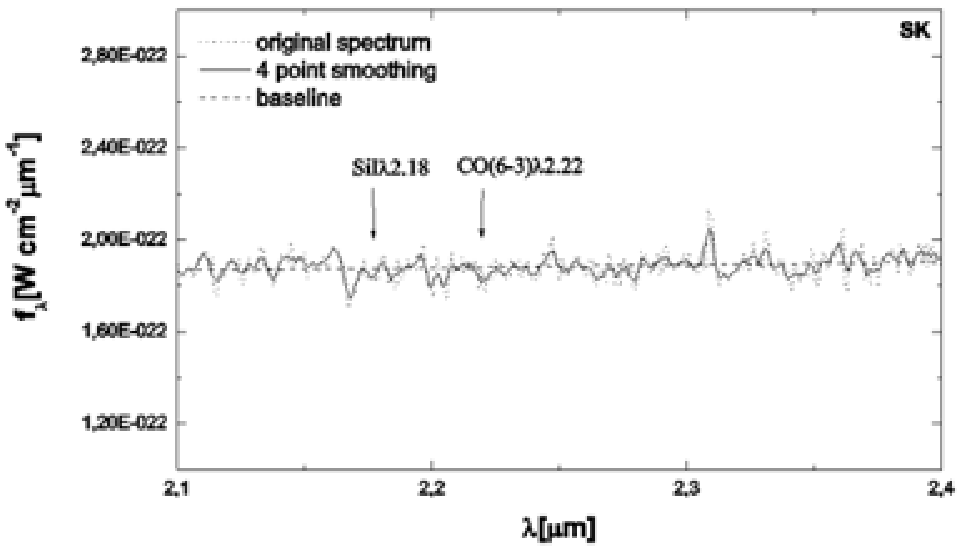}}
\caption{$K$ band spectrum of the nuclear region of 3C 48. The dotted line is the original spectrum, whereas the solid line shows the spectrum after a 4 point smoothing. Indicated are the redshifted positions of the \ion{Si}{I} and CO(6-3) absorption lines.
}
\label{3c48SKSpec}
\end{figure}

The $K$ spectrum (Fig.~\ref{3c48SKSpec}) is less noisy but there is
hardly any feature to be recognized at its original spectral
resolution of $\Delta\lambda=0.005\ \mu$m. Therefore we smoothed the
spectrum with a boxcar of a width of 4 pixels. There are two
features appearing in the spectrum, one left of the indicated \ion{Si}{I} line
and one at a wavelength of about 2.3~$\mu$m. The latter one could not
be identified. At this location there seems to be no prominent
emission line. We suggest this to be a noise peak. The absorption
feature shortward of the \ion{Si}{I} line is most probably an atmospheric or
stellar residual from the telluric-line removal and can therefore not
be attributed to 3C 48. The noise per pixel of the smoothed spectrum
is about half of the noise in the original spectrum
\begin{equation}
\sigma_\mathrm{pixel}=3\times 10^{-24}\ \mathrm{W~cm}^{-2}\ \mu\mathrm{m}^{-1}.
\end{equation}

We investigated the noise features at the CO(6-3) line at a
redshifted wavelength of 2.217~$\mu$m (cf. Table
\ref{3C48DiagnosticLines}). The CO(6-3)~$\lambda 1.6185$ line is a
tracer of the stellar content of a galaxy, as this line arises in the
atmospheres of cool stars \citep{1993A&A...280..536O}. It is weak in
early K- but strong in late M-type stars. We determined the line flux
by using the triangular area above the absorption line. The flux and
noise integrated over a wavelength interval of
$\Delta\lambda=10$~nm is
\begin{eqnarray}
f_\mathrm{CO(6-3)} &\le& 2.43\times 10^{-26}\ \mathrm{W~cm}^{-2} \\
\sigma_\mathrm{integrated} &=& 2.46\times 10^{-26}\ \mathrm{W~cm}^{-2}\nonumber\\
\end{eqnarray}
where the noise was determined by measuring the fluctuations
throughout the spectrum except at the CO(6-3) wavelength.
We thus have a $1\sigma$ detection of the line. The depth of the line
is about 3\% of the continuum level at that wavelength. According to
\cite{1993A&A...280..536O} the typical depth of this line is
about 15\% of the continuum for a pure stellar contribution. Taking
into account our photometrically determined stellar contribution of about 30\%
of the galaxy light, this leads to an expected CO(6-3) line depth of
$\sim 4\%$. Accordingly our spectroscopic result seems to be
consistent with the photometric results within the errors.

\section{Summary}
We have presented the most recent imaging and spectroscopy data on 3C
48 obtained with ISAAC at the ESO VLT UT1 (Antu). Based on these
sensitive high angular resolution data for the first time we can
identify the disputed secondary nucleus 3C 48A in the NIR. The location of
3C 48A in the $J-H/H-K$ diagram indicates the presence of
warm dust, probably heated by starbursts or ongoing starformation discussed in  \citet{2000ApJ...528..201C}. Our NIR images do not allow us to decide whether 3C  48A is a true second nucleus (e.g. of a merging companion) or rather due to a jet-ISM interaction.

We have shown that the NIR colors of 3C 48 are reddened compared to a
pure 10~Gyr-old stellar population and therefore extinction due to
molecular gas and dust plays a role throughout the host galaxy. 
Measurement of the colors in the NW tidal tail indicate a younger (blue
$J-H$) but dust-associated (red $H-K$) stellar population. However, a
high visual extinction seems to be at work as un-reddened young main
sequence populations are located further to the blue in
Fig. \ref{3c48TwoColors}. Therefore deep, high resolution NIR
spectroscopy of the host galaxy is needed.

The QSO-subtracted colors compared to pure 10~Gyr stellar colors show a
stellar contribution to the overall galaxy continuum light of about
30\% in $Ks$, which is consistent with the 1$\sigma$ depth limit of the
CO(6-3) absorption line in the $K$-band spectrum.

We can see tidal features in the images obtained in all three
NIR bands. Judging from the colors, the apparent counter-tidal tail to
the southeast of the QSO more likely seems to be an un-associated
galaxy in the foreground, whereas \cite{2000ApJ...528..201C} find it
to be a background galaxy at a redshift of $z$=0.8. 
Despite this discrepancy it seems to be established that the apparent
counter-tidal tail is not associated with 3C 48. Therefore further
investigation is necessary. Another solution to the missing counter
tidal tail is given by multi particle simulations \citep{scharw}, which
show that there could be a counter tidal tail in a SW-NE
direction in front of the galaxy. This is in correspondence with the
velocity dispersions measured by \cite{2000ApJ...528..201C} in the
optical along their slit B.

Its large host galaxy which is clearly subject to strong tidal
interactions makes 3C 48 an ideal candidate for further studies of the
merger-induced starformation at its redshift of $z=0.37$.

\begin{acknowledgements}
This work was supported in part by the Deutsche Forschungsgemeinschaft
(DFG) via grant SFB 494.
We kindly thank T. Ott, MPE Garching, for interest and initial support. We also kindly thank P. N. Wilkinson for the permission to use the 1.66 GHz continuum map.

\end{acknowledgements}

\bibliographystyle{aa}
\bibliography{H4724}

\begin{thebibliography}{24}
\expandafter\ifx\csname natexlab\endcsname\relax\def\natexlab#1{#1}\fi

\bibitem[{{Boroson} \& {Oke}(1984)}]{1984ApJ...281..535B}
{Boroson}, T.~A. \& {Oke}, J.~B. 1984, \apj, 281, 535

\bibitem[{{Boyce} {et~al.}(1999){Boyce}, {Disney}, \&
  {Bleaken}}]{1999MNRAS.302L..39B}
{Boyce}, P.~J., {Disney}, M.~J., \& {Bleaken}, D.~G. 1999, \mnras, 302, L39

\bibitem[{{Bruzual A.} \& {Charlot}(1993)}]{1993ApJ...405..538B}
{Bruzual A.}, G. \& {Charlot}, S. 1993, \apj, 405, 538

\bibitem[{{Canalizo} \& {Stockton}(2000)}]{2000ApJ...528..201C}
{Canalizo}, G. \& {Stockton}, A. 2000, \apj, 528, 201

\bibitem[{{Chatzichristou} {et~al.}(1999){Chatzichristou}, {Vanderriest}, \&
  {Jaffe}}]{1999A&A...343..407C}
{Chatzichristou}, E.~T., {Vanderriest}, C., \& {Jaffe}, W. 1999, \aap, 343, 407

\bibitem[{{Frogel} {et~al.}(1978){Frogel}, {Persson}, {Matthews}, \&
  {Aaronson}}]{1978ApJ...220...75F}
{Frogel}, J.~A., {Persson}, S.~E., {Matthews}, K., \& {Aaronson}, M. 1978,
  \apj, 220, 75

\bibitem[{{Glass}(1999)}]{1999hia..book.....G}
{Glass}, I.~S. 1999, {Handbook of infrared astronomy} (Cambridge University
  Press)

\bibitem[{{Glass} \& {Moorwood}(1985)}]{1985MNRAS.214..429G}
{Glass}, I.~S. \& {Moorwood}, A.~F.~M. 1985, \mnras, 214, 429

\bibitem[{{Hawarden} {et~al.}(2001){Hawarden}, {Leggett}, {Letawsky},
  {Ballantyne}, \& {Casali}}]{2001MNRAS.325..563H}
{Hawarden}, T.~G., {Leggett}, S.~K., {Letawsky}, M.~B., {Ballantyne}, D.~R., \&
  {Casali}, M.~M. 2001, \mnras, 325, 563

\bibitem[{{Hutchings} \& {Neff}(1997)}]{1997AJ....113..550H}
{Hutchings}, J.~B. \& {Neff}, S.~G. 1997, \aj, 113, 550

\bibitem[{{Moorwood}(1995)}]{1995SPIE.2475..262M}
{Moorwood}, A.~F. 1995, in Proc. SPIE Vol. 2475, p. 262-267, Infrared Detectors
  and Instrumentation for Astronomy, Albert M. Fowler; Ed., Vol. 2475, 262--267

\bibitem[{{Neugebauer} {et~al.}(1985){Neugebauer}, {Soifer}, \&
  {Miley}}]{1985ApJ...295L..27N}
{Neugebauer}, G., {Soifer}, B.~T., \& {Miley}, G.~K. 1985, \apjl, 295, L27

\bibitem[{{Origlia} {et~al.}(1993){Origlia}, {Moorwood}, \&
  {Oliva}}]{1993A&A...280..536O}
{Origlia}, L., {Moorwood}, A.~F.~M., \& {Oliva}, E. 1993, \aap, 280, 536

\bibitem[{{Osterbrock}(1989)}]{osterbrock89:_astrop_gaseous_nebul_activ_galac_%
nuclei}
{Osterbrock}, D.~E. 1989, Astrophysics of Gaseous Nebulae and Active Galactic
  Nuclei (University Science Books)

\bibitem[{{Persson} {et~al.}(1998){Persson}, {Murphy}, {Krzeminski}, {Roth}, \&
  {Rieke}}]{1998AJ....116.2475P}
{Persson}, S.~E., {Murphy}, D.~C., {Krzeminski}, W., {Roth}, M., \& {Rieke},
  M.~J. 1998, \aj, 116, 2475

\bibitem[{{Rieke} \& {Lebofsky}(1985)}]{1985ApJ...288..618R}
{Rieke}, G.~H. \& {Lebofsky}, M.~J. 1985, \apj, 288, 618

\bibitem[{{Sanders} {et~al.}(1988){Sanders}, {Soifer}, {Elias}, {Madore},
  {Matthews}, {Neugebauer}, \& {Scoville}}]{1988ApJ...325...74S}
{Sanders}, D.~B., {Soifer}, B.~T., {Elias}, J.~H., {et~al.} 1988, \apj, 325, 74

\bibitem[{{Scharw\"achter} {et~al.}(2003){Scharw\"achter}, {Eckart},
  {Pfalzner}, {Zuther}, {Krips}, \& C.}]{scharw}
{Scharw\"achter}, J., {Eckart}, A., {Pfalzner}, S., {et~al.} 2003, accepted
  \aap, astro-ph/0310740

\bibitem[{{Scoville} {et~al.}(1993){Scoville}, {Padin}, {Sanders}, {Soifer}, \&
  {Yun}}]{1993ApJ...415L..75S}
{Scoville}, N.~Z., {Padin}, S., {Sanders}, D.~B., {Soifer}, B.~T., \& {Yun},
  M.~S. 1993, \apjl, 415, L75

\bibitem[{{Stockton} \& {Ridgway}(1991)}]{1991AJ....102..488S}
{Stockton}, A. \& {Ridgway}, S.~E. 1991, \aj, 102, 488

\bibitem[{{Toomre} \& {Toomre}(1972)}]{1972ApJ...178..623T}
{Toomre}, A. \& {Toomre}, J. 1972, \apj, 178, 623

\bibitem[{{Tyul'bashev} \& {Chernikov}(2001)}]{2001A&A...373..381T}
{Tyul'bashev}, S.~A. \& {Chernikov}, P.~A. 2001, \aap, 373, 381

\bibitem[{{Wilkinson} {et~al.}(1991){Wilkinson}, {Tzioumis}, {Benson},
  {Walker}, {Simon}, \& {Kahn}}]{1991Natur.352..313W}
{Wilkinson}, P.~N., {Tzioumis}, A.~K., {Benson}, J.~M., {et~al.} 1991, \nat,
  352, 313

\bibitem[{{Wink} {et~al.}(1997){Wink}, {Guilloteau}, \&
  {Wilson}}]{1997A&A...322..427W}
{Wink}, J.~E., {Guilloteau}, S., \& {Wilson}, T.~L. 1997, \aap, 322, 427

\end{thebibliography}
\end{document}